\documentclass{article}
\usepackage{latexsym}
\usepackage{amsfonts}
\usepackage{amssymb}
\usepackage{amsmath}
\def\abstract#1{\vskip 7mm 
        \begin{center}{\large Abstract}\par \smallskip
                \begin{minipage}[c]{12cm}
                        \small #1
                \end{minipage}
        \end{center}
}
\def\title#1{\begin{center}{\Large\bf #1}\end{center}}
\def\author#1{\vskip 5mm \begin{center}{#1}\end{center}}
\def\address#1{\begin{center}{\it #1}\end{center}}
\makeatletter
\@ifundefined{lesssim}{}{}
\@ifundefined{gtrsim}{}{}
\def\vereq#1#2{\lower3pt\vbox{\baselineskip1.5pt \lineskip1.5pt
\ialign{$\m@th#1\hfill##\hfil$\crcr#2\crcr\sim\crcr}}}
\makeatother

\begin{document}

\title{%
Gravitational Waves from Rotating Black Strings }

\author{%
Sugumi Kanno\footnote{E-mail:sugumi@tap.scphys.kyoto-u.ac.jp},
Jiro Soda\footnote{E-mail:jiro@tap.scphys.kyoto-u.ac.jp}
}
\address{%
  Department of Physics,
  Kyoto University, Kyoto 606-8501, Japan
}

\abstract{
  In the Randall-Sundrum two-brane model (RS1), a Kerr black hole on the
 brane can be naturally identified with a section of rotating 
  black string.  To estimate  Kaluza-Klein (KK) corrections on 
 gravitational waves emitted by perturbed rotating black strings, we give 
  the effective Teukolsky equation on the brane which is separable
  equation and hence numerically manageable. In this process, we derive
  the master equation for the electric part of the Weyl tensor $E_{\mu\nu}$
  which would be also useful to discuss the transition from black strings
  to localized black holes triggered by Gregory-Laflamme instability.   
}

\section{Introduction}

  The recent progress of superstring theory has provided a new picture of
  our universe, the so-called braneworld.
  The evidence of the extra dimensions in this scenario
  should be explored in the early stage of the universe
 or the  black hole. In particular,  gravitational waves are key probes 
  because they can propagate into the bulk freely. 
  Cosmology in this scenario has been investigated
 intensively. While, gravitational waves from black holes have been
  less studied so far. In this paper, we shall take a step 
  toward this direction.    
 
  Here, we will concentrate on a two-brane model which is proposed 
  by Randall and Sundrum as a simple and 
  phenomenologically interesting model~\cite{RS}.
 In this RS1 model, the large black hole on the brane 
 is expected to be black string.  Hence, it would be important
 to clarify how the gravitational waves are generated in the perturbed 
  black string system and how the effects of the extra dimensions come 
  into the observed signal of the gravitational waves. 
 It is desired to have a basic formalism for analyzing gravitational 
 waves generated by perturbed rotating black string. 
  
 It is well known in general relativity that the perturbation around Kerr black 
 hole is elegantly treated in the Newman-Penrose formalism.
 Indeed, Teukolsky derived a separable master equation for the gravitational
 waves in the Kerr black hole background~\cite{Teukolsky}. 
  The main purpose of this paper is to extend the Teukolsky
 formalism to the braneworld context and derive the effective Teukolsky
 equation.  
  
  The organization of this paper is as follows.
  In sec.II, we present the model and 
  demonstrate the necessity of solving $E_{\mu\nu}$ in deriving
   the effective Teukolsky equation. 
   In sec.III, a perturbed equation and the junction conditions
   for $  E_{\mu\nu}$ are obtained.
  In sec.IV, we give the explicit solution for $ E_{\mu\nu}$  
  using the gradient expansion method. Then, the effective Teukolsky
  equation is presented. 
  The final section is devoted to the conclusion.


\section{ Teukolsky Equation on  the Brane}

Based on the Newmann-Penrose (NP) null-tetrad formalism, in which the
tetrad components of the curvature tensor are the fundamental variables,
a master equation for the curvature perturbation was developed by Teukolsky
for a Kerr black hole with source. The master equation is called the Teukolsky
equation, and it is a wave equation for a null-tetrad component of the 
Weyl tensor $\Psi_0=-C_{pqrs}\ell^pm^q\ell^rm^s$ or $\Psi_4=-C_{pqrs}n^p\bar{m}^qn^r\bar{m}^s$, where $C_{pqrs}$ is the Weyl tensor and $\ell, n, m, \bar{m}$ 
are null basis in the NP formalism. All information about the 
gravitational radiation flux at infinity and at the event horizon can be 
extracted from $\Psi_0$ and $\Psi_4$. 
The Teukolsky equation is constructed by combining the Bianchi identity
with the Einstein equations. The Riemann tensor in the Bianchi identity
is written in terms of the Weyl tensor and the Ricci tensor. The Ricci
tensor is replaced by the matter fields using the Einstein equations.
In this way, the Bianchi identity becomes no longer identity and one
can get a master equation in which the curvature tensor is the fundamental 
variable.

It is of interest to consider the four-dimensional effective Teukolsky
equation in braneworld to investigate  gravitational
waves from perturbed rotating black strings. 
We consider an $S_1/Z_2$ orbifold space-time with the two branes 
as the fixed points. In this RS1 model, 
the two  3-branes are embedded in AdS$_5$ with the curvature
radius $\ell$ and the brane tensions given by
$\sigma_\oplus=6/(\kappa^2\ell)$ and 
$\sigma_\ominus=-6/(\kappa^2\ell)$. 
Our system is described by the action 
\begin{eqnarray}
S=\frac{1}{2 \kappa^2}\int d^5 x 
	\sqrt{-\overset{(5)}{g}}\left({\cal R}
	+\frac{12}{\ell^2}\right)
	-\sum_{i=\oplus,\ominus}\sigma_i 
	\int d^4 x \sqrt{-g^{i\mathrm{\hbox{-}brane}}}
	+\sum_{i=\oplus,\ominus} 
	\int d^4 x \sqrt{-g^{i\mathrm{\hbox{-}brane}}}
	\,{\cal L}_{\rm matter}^i \ ,\label{5D:action}
\end{eqnarray}
where $\overset{(5)}{g}_{\mu\nu}$, ${\cal R}$, 
$g^{i\mathrm{\hbox{-}brane}}_{\mu\nu}$, and $\kappa^2$ 
are the 5-dimensional metric, the 5-dimensional scalar curvature, 
the induced metric on the $i$-brane, 
and the 5-dimensional gravitational constant, respectively.

Since the Bianchi identity is
independent of dimensions, what we need is the projected Einstein equations
on the brane derived by Shiromizu, Maeda and Sasaki~\cite{ShiMaSa}. 
The first order perturbation of the projected Einstein equation is 
\begin{eqnarray}
G_{\mu\nu}=8\pi GT_{\mu\nu}- \delta E_{\mu\nu} \ ,
\label{SMS}
\end{eqnarray}
where $8\pi G =\kappa^2 /\ell$.
If we replace the Ricci curvature in the Bianchi identity to
the matter fields and $E_{\mu\nu}$ using Eq.~(\ref{SMS}), then the projected 
Teukolsky equation on the brane is written in the following form,
\begin{eqnarray}
&&[(\triangle+3\gamma-\gamma^{\ast}+4\mu+\mu^{\ast})
	(D+4\epsilon-\rho)
	-(\delta^{\ast}-\tau^{\ast}+\beta^{\ast}+3\alpha+4\pi)
	(\delta-\tau+4\beta)-3 \Psi_2]\delta \Psi_4 \nonumber\\
&&\hspace{1cm}
	={1\over 2} (\triangle+3\gamma-\gamma^{\ast}+4\mu+\mu^{\ast})
	\nonumber \\
&&\hspace{1cm}
	\times[(\delta^{\ast}-2\tau^{\ast}+2\alpha)
	(8\pi G T_{nm^{\ast}}- \delta E_{nm^{\ast}})
	-(\triangle+2\gamma-2\gamma^{\ast}+\mu^{\ast})
	(8\pi G T_{m^{\ast}m^{\ast}}
	- \delta E_{m^{\ast}m^{\ast}})]\nonumber\\
&&\hspace{1cm}
	+{1\over 2}(\delta^{\ast}-\tau^{\ast}+\beta^{\ast}+3\alpha+4\pi)
	\nonumber \\
&&\hspace{1cm}
	\times[(\triangle+2\gamma+2\mu^{\ast})(8\pi G T_{nm^{\ast}}
	- \delta E_{nm^{\ast}})
	-(\delta^{\ast}-\tau^{\ast}+2\beta^{\ast}+2\alpha)
	(8\pi G T_{nn}- \delta E_{nn})] \ .
	\label{teukolsky}
\end{eqnarray}
Here our notation follows that of  \cite{Teukolsky}.
We see the effects of a fifth dimension, $\delta E_{\mu\nu}$, is described as
a source term in the projected Teukolsky equation. It should be stressed that
 the projected Teukolsky equation on the brane Eq.~(\ref{teukolsky})
is not a closed system yet. One must solve the gravitational field in the bulk
 to obtain $\delta E_{\mu\nu}$. 


\section{Master Equation for $\delta E_{\mu\nu}$}

 Starting with the 5-dimensional Bianchi identities, 
 we can derive the perturbed equation for $\delta E_{\mu\nu}$. 
The background we consider is a Ricci flat string without source 
($T_{\mu\nu}=0$) whose metric is written as
\begin{eqnarray}
ds^2=dy^2+e^{-2\frac{y}{\ell}}g_{\mu\nu}(x^\mu)dx^\mu dx^\nu \ ,
\end{eqnarray}
where $g_{\mu\nu}(x^\mu)$ is supposed to be the Ricci flat metric. 
 The equation of motion for $\delta E_{\mu\nu}$ in the bulk is found as
\begin{eqnarray}
\left(\partial^2_y-\frac{4}{\ell}\partial_{y}+\frac{4}{\ell^2}\right)
	\delta E_{\mu\nu}&=& 
	-e^{2\frac{y}{\ell}}{\hat{\cal L}_{\mu\nu}{}^{\alpha\beta}}
	\delta E_{\alpha\beta} 
     \equiv - e^{2\frac{y}{\ell}}\hat{\cal L}\delta E_{\mu\nu} \ ,
	\label{eom:weyl}
\end{eqnarray}
where ${\hat{\cal L}_{\mu\nu}{}^{\alpha\beta}}$
stands for the Lichnerowicz operator,
\begin{eqnarray}
{\hat{\cal L}_{\mu\nu}{}^{\alpha\beta}}
	= \Box \delta_\mu^\alpha \delta_\nu^\beta 
	+2 R_\mu{}^\alpha{}_\nu{}^\beta  \ .
\end{eqnarray}
Here, the covariant derivative and the Riemann tensor are constructed 
 from $g_{\mu\nu} (x)$. 

 On the other hand, the junction conditions on each branes become
\begin{eqnarray}
&& e^{2\frac{y}{\ell}}\left[e^{-2\frac{y}{\ell}}\delta E_{\mu\nu}\right]_{,y}
\bigg{|}_{y=0}  =-\frac{\kappa^2}{6}\overset{\oplus}{T}_{|\mu\nu}
	-\frac{\kappa^2}{2}{\hat{\cal L}_{\mu\nu}{}^{\alpha\beta}}
	\left( \overset{\oplus}{T}_{\alpha\beta}
	-\frac{1}{3}g_{\alpha\beta}\overset{\oplus}{T}\right) \ ,
	\label{JC:p-1}\ \\
&&e^{2\frac{y}{\ell}}\left[e^{-2\frac{y}{\ell}}\delta E_{\mu\nu}\right]_{,y}
\bigg{|}_{y=d} =\frac{\kappa^2}{6}\overset{\ominus}{T}_{|\mu\nu}
	+\frac{\kappa^2}{2}{\hat{\cal L}_{\mu\nu}{}^{\alpha\beta}}
	\left( \overset{\ominus}{T}_{\alpha\beta}
	-\frac{1}{3}g_{\alpha\beta}\overset{\ominus}{T}\right) \ .
	\label{JC:n-1}
\end{eqnarray}
Let $\overset{\oplus}{\phi}(x)$ and $\overset{\ominus}{\phi}(x)$
be the scalar fields on each branes which satisfy
\begin{eqnarray}
\Box \overset{\oplus}{\phi}={\kappa^2\over 6} 
	\overset{\oplus}{T} \ , \quad
\Box \overset{\ominus}{\phi}={\kappa^2\over 6} 
	\overset{\ominus}{T} \ ,
	\label{radion2}
\end{eqnarray}
respectively. In the Ricci flat space-time, the identity
\begin{eqnarray}
\left(\Box\phi\right)_{|\mu\nu}
	={\hat{\cal L}_{\mu\nu}{}^{\alpha\beta}}
	\left(\phi_{|\alpha\beta}\right)
	\label{relation}
\end{eqnarray}
holds. Thus, Eqs.(\ref{JC:p-1}) and (\ref{JC:n-1}) can be rewritten as
\begin{eqnarray}
&&\hspace{-5mm}
e^{2\frac{y}{\ell}}
	\left[
	e^{-2\frac{y}{\ell}}\delta E_{\mu\nu}
	\right]_{,y}
	\bigg{|}_{y=0} 
	=- \hat{\cal L} \left(\overset{\oplus}{\phi}_{|\mu\nu}
	- g_{\mu\nu} \Box \overset{\oplus}{\phi}\right)
	-\frac{\kappa^2}{2} \hat{\cal L}
	 \overset{\oplus}{T}_{\mu\nu}
	~\equiv \hat{\cal L} \overset{\oplus}{S}_{\mu\nu} \ ,
	\label{JC:p}\ \\
&&\hspace{-5mm}
e^{2\frac{y}{\ell}}
	\left[
	e^{-2\frac{y}{\ell}}\delta E_{\mu\nu}
	\right]_{,y}
	\bigg{|}_{y=d} 
	=\hat{\cal L}\left(\overset{\ominus}{\phi}_{|\alpha\beta}
	- g_{\mu\nu} \Box \overset{\ominus}{\phi} \right)
	+\frac{\kappa^2}{2} \hat{\cal L}
	 \overset{\ominus}{T}_{\mu\nu}
	~\equiv - \hat{\cal L} \overset{\ominus}{S}_{\mu\nu} \ .
	\label{JC:n}
\end{eqnarray}
The scalar fields $\overset{\oplus}{\phi}$ and $\overset{\ominus}{\phi}$
 can be interpreted as the brane fluctuation modes.

\section{Effective Teukolsky Equation}

Now, we solve Eq.(\ref{eom:weyl}) under the boundary conditions
(\ref{JC:p}) and (\ref{JC:n}) using the gradient expansion method.

\subsection{Gradient Expansion Method}

 It is known that the Gregory-Laflamme instability occurs if 
 the curvature length scale of the black hole $L$ is less than the Compton
 wavelength of KK modes $\sim \ell \exp(d/\ell) $.  
 As we are interested in the stable rotating black string, we assume
\begin{equation}
\epsilon~=~ \left({\ell \over L}\right)^2 \ll 1  \ .
\end{equation}
 This means that the curvature on the brane can be neglected 
compared with the derivative with respect to $y$. 
 Our iteration scheme consists in writing the Weyl tensor $E_{\mu\nu}$
 in  the order of $\epsilon$~\cite{Kanno}. 
Hence, we will seek the Weyl tensor as a perturbative series
\begin{eqnarray}
\delta E_{\mu\nu}(y,x^\mu ) 
 =  \delta\overset{(1)}{E}_{\mu\nu} (y,x^\mu)
	+\delta\overset{(2)}{E}_{\mu\nu}(y, x^\mu )
	+\delta\overset{(3)}{E}_{\mu\nu}(y, x^\mu ) + \cdots  \ .
	\label{series}
\end{eqnarray}
Substituting this Eq.(\ref{series}) into Eq.(\ref{eom:weyl}),
 we can obtain $\delta E_{\mu\nu}$ perturbatively. 

\subsection{Effective Teukolsky Equation}

 As a result, Teukolsky equation  takes the following form
\begin{eqnarray}
    \hat{P} \delta \Psi_4 
    = \hat{Q} \left( 8\pi G T_{nm^{\ast}} - \delta E_{nm^{\ast}} \right) 
    + \cdots \ ,
\end{eqnarray}
where $\hat{P}$ and $\hat{Q}$ are operators defined in (\ref{teukolsky}). 
 What we needed is $\delta E_{\mu\nu}$ in the above equation. 
 We can write down $\delta E_{\mu\nu}$ on the brane up to the second order
 as~\cite{Kanno:2003au}
\begin{eqnarray}
 \delta E_{\mu\nu}\Big|_{y=0} &=& {2\over \ell} {\Omega^2 \over 1-\Omega^2} 
	\left[  \overset{\oplus}{S}{}_{\mu\nu} 
	+ \Omega^2  \overset{\ominus}{S}
	{}_{\mu\nu} \right] \nonumber \\
&&	+ \left[ {1\over 2} + {d\over \ell} {1\over \Omega^2 -1} \right]
	{\ell \over 1-\Omega^2} 
	\left[ \hat{\cal L} \overset{\oplus}{S}{}_{\mu\nu} 
	+ \Omega^4 \hat{\cal L} \overset{\ominus}{S}
	{}_{\mu\nu} \right]
	+  {\ell \over 4} 
	\left[ \hat{\cal L} \overset{\oplus}{S}{}_{\mu\nu} 
	+ \Omega^2 \hat{\cal L} \overset{\ominus}{S}
	{}_{\mu\nu} \right] 
	\label{extrasource}\ ,
\end{eqnarray}
where
\begin{eqnarray}
\overset{\oplus}{S}_{\mu\nu}
    &=&   - \overset{\oplus}{\phi}_{|\mu\nu}
         + g_{\mu\nu} \Box \overset{\oplus}{\phi}
	-\frac{\kappa^2}{2} \overset{\oplus}{T}_{\mu\nu}
	\ , 
                     \\
\overset{\ominus}{S}_{\mu\nu}
    &=& - \overset{\ominus}{\phi}_{|\mu\nu}
          + g_{\mu\nu} \Box \overset{\ominus}{\phi}
	-\frac{\kappa^2}{2} \overset{\ominus}{T}_{\mu\nu}
	\ . 
\end{eqnarray}
Substituting this $\delta E_{\mu\nu}$ into (\ref{teukolsky}), 
we get the effective
 Teukolsky equation on the brane. 
  From Eq.~(\ref{extrasource}), we see KK corrections give extra sources
 to Teukolsky equation. To obtain quantitative results, we must resort
  to numerical calculations. It should be stressed that the effective
 Teukolsky equation is separable like as the conventional Teukolsky equation.
 Therefore, it is suitable for numerical treatment. 
 
 Notice that our result in this section  assume only Ricci flatness. 
 If we do not care about separability, we can study the gravitational waves
 in the general Ricci flat background. 
  Moreover, we can analyze other types of waves using $\delta E_{\mu\nu}$. 
 As $\delta E_{\mu\nu}$ has 5 degrees of freedom
 which corresponds to the degrees of freedom of the bulk gravitational waves,
 one expect the scalar gravitational waves and vector gravitational waves.
 Without KK effects, no vector gravitational waves exist and the scalar
 gravitational waves can be described as the Brans-Dicke scalar waves. 
  However, KK effects produce new effects which might be observable.

\section{Conclusion}

We  formulated the perturbative formalism around the Ricci
 flat two-brane system. In particular, the master equation for 
 $\delta E_{\mu\nu}$ is derived. 
  The gradient expansion method is utilized to get
 a series solution. This gives the closed system of equations
 which we call the effective Teukolsky equations in the case of type D
 induced metric on the brane. 
 This can be used for estimating the Kaluza-Klein corrections on 
 the gravitational waves emitted from the perturbed rotating black string.
 Our effective Teukolsky equation is completely separable, hence the
 numerical scheme can be developed in a similar manner as was done 
 in the case of 4-dimensional Teukolsky equation.

\smallskip
\noindent{Acknowledgements}\\
This work was supported in part by  Grant-in-Aid for  Scientific
Research Fund of the Ministry of Education, Science and Culture of Japan 
 No. 155476 (SK) and  No.14540258 (JS) and also
  by a Grant-in-Aid for the 21st Century COE ``Center for
  Diversity and Universality in Physics".

\end{document}